\documentstyle[epsf]{EuroPhys}
\input EuroMacr

\newcommand{\rf}[1]{(\ref{#1})}
\newcommand{\AC}{{\cal A}}

\newcommand{\tk}{\tilde k}
\newcommand{\del}{\delta}
\newcommand{\vep}{\varepsilon}
\newcommand{\th}{\theta}

\newcommand{\om}{\omega}

\newcommand{\al}{\alpha}
\newcommand{\sg}{\sigma}
\newcommand{\be}{\begin{equation}}
\newcommand{\bh}{\begin{mathletters}}
\newcommand{\eh}{\end{mathletters}}
\newcommand{\ee}{\end{equation}}
\newcommand{\ba}{\begin{eqnarray}}

\newcommand{\ea}{\end{eqnarray}}
\newcommand{\br}{\begin{array}}
\newcommand{\er}{\end{array}}
\newcommand{\dx}[2]{\partial_{#2}{#1}}
\newcommand{\ddx}[2]{\partial^2_{#2}{#1}}

\begin{document}
\euro{}{}{}{}
\Date{}

\shorttitle{I.KELLER \etal ENVELOPE DYNAMICS IN 2D FARADAY WAVE}

 \title{On  envelope dynamics in 2D Faraday waves}
\author{ Igor Keller, Alexander Oron, Pinhas Z. Bar-Yoseph }
\institute{  Faculty of Mechanical
Engineering,\\  Technion - Israel Institute of Technology, Haifa 32000, Israel}

\rec{}{}
\pacs{
%\Pacs{47}{54$+$r}{Riemannian geometries}
\Pacs{47}{35$+$i}{Hydrodynamic waves}
\Pacs{47}{20$-$k}{Hydrodynamic stability}
}
\maketitle

\abstract{A weakly nonlinear   model for two-dimensional Faraday
 waves over infinite depth  is derived and studied. 
Sideband instability of monochromatic standing waves as well 
as non-monochromatic solutions are studied analytically.
Persistent irregular regimes  are found numerically.
    }

\section{ Introduction}
 Driven gravity-capillary waves is a well-known pattern-forming system owing its 
 popularity to a relative  simplicity  and diversity  of observed phenomena. 
Despite  the wide interest in the problem, the theoretical basis of  pattern 
formation  in large-aspect-ratio  systems is not yet fully understood.
According to experimental data \cite{Chris,Ezerskii} the complex dynamics 
are observed  even  at small supercriticalities. 
In the derivation of a theoretical model one  must, therefore,  consistently 
incorporate the effect of viscous dissipation, which governs  the stability 
threshold and supercritical regimes.
 Furthermore, description of   spatially non-homogeneous pattern formation 
involves  ``slow'' space variables, whose scalings  in a weakly supercritical domain 
depend strongly  on the way how the dissipation was taken into account. 

Some of earlier  models   \cite{Ezerskii,Miles}  describing the  envelope dynamics of 
driven waves in a large-aspect-ratio system
have been  derived from the equations for an inviscid fluid 
with damping  incorporated  phenomenologically 
 by  adding linear damping terms to the amplitude equations.
As a result, an amplitude saturation due to  nonlinear damping, which 
remains important  even for  low viscosity \cite{We}, is thus overlooked.
In addition, the model obtained  in \cite{Miles} does not possess the symmetries of the 
underlying  physical system.
The nonlinear damping terms are included in  some other models \cite{Milner,Cherep}. 
However, due to neglecting of the  boundary layer in the derivation of \cite{Milner},
 those terms are of higher asymptotic order than in our derivation (see below).
In spite of using the   consistent multiple-scales method    \cite{Cherep}
with  the viscous boundary layer rigorously accounted for,  due to the
  chosen  scalings the validity of the model  seems to be limited to 
exceedingly small   supercriticalities.   

In this letter we derive using results of \cite{We}  
a single amplitude equation describing the evolution of 
one-dimensional patterns for  small supercriticalities.
This evolution equation is further investigated analytically 
and numerically.

\section{Model equations} In a low-viscosity  deep-water  large-aspect-ratio  system
a packet of  modes with adjacent wavenumbers  becomes excited due to the 
parametric subharmonic resonance and   locked to the 
(weak)  forcing. 
In this case the principal part of  a 2D motion of the free surface 
 can be represented as  a superposition of two  counter-propagating small 
and narrow wave packets      
\be
\zeta(x,t) = \hat ae^{i\tk x+ i\om t}+
\hat be^{i\tk x - i\om t}+c.c.
\ee
Here $\om$ is  half  the  driving frequency  and the corresponding  wavenumber 
$\tk$ is obtained from  the dispersion relation $\om^2=g\tk+(\Sigma/\rho) \tk^3$, where 
$\Sigma$ is  surface tension, $g$ is the gravity acceleration and  $\rho$ is
 the fluid density. 

In the case of  weak  forcing  and low viscosity  
the   slow   spatio-temporal evolution of the envelopes 
 $\hat a(x,t),\hat b(x,t)$ is 
described by the set of equations \cite{We}
\ba
  \dx{\hat a}{t}- s  \dx{\hat a}{x}=-(\gamma-\frac12\gamma^{\frac32})\hat a+ i f  \hat b +(ic_1-p_1\sqrt{\gamma}) |\hat a|^2\hat a +
(ic_2-p_2\sqrt{\gamma})|\hat b|^2 \hat a,\label{ampl_eq_a}
\\
\dx{\hat b}{t}+ s \dx{\hat b}{x}= -(\gamma-\frac12\gamma^{\frac32})\hat b - i f  \hat a -(ic_1+p_1^*\sqrt{\gamma})|\hat b|^2\hat b - 
(ic_2+p_2^*\sqrt{\gamma})|\hat a|^2 \hat b,\label{ampl_eq_b}
\ea
where an asterisk denotes complex conjugate.
 The coefficients  in eqs.\rf{ampl_eq_a},\rf{ampl_eq_b} are: 
 a finite group speed $s=d\om(\tk)/dk$, a small  linear damping $\gamma$ and 
the forcing amplitude  $f$,  and real-valued  nonlinear dispersion coefficients, 
$c_1,c_2$,  given by
$$
 s=\sg+\frac12, \quad
\gamma=\frac{2\nu\tk^2}{\om},\quad f=\frac{\hat f}4(1-\sg),\quad
\quad\sg=\frac{\Sigma \tk^3}{\rho \om^2},
\quad c_1=\frac34\sg-\frac{3\sg-2}{3\sg-1},\quad
c_2=\frac32\sg+\frac{6\sg+4}{3\sg+1}. 
$$
Here $\hat f$  is the amplitude of forcing (in units of $g$),
 $\sg$ describes the nature of the wave: it is a gravity wave or a
 capillary wave in two  limiting cases of 
$\sg=0$ and $\sg=1$, respectively. 

Eqs.\rf{ampl_eq_a},\rf{ampl_eq_b} were derived 
in \cite{We} from the Navier-Stokes equations 
using the multiple-scales  method.
 Scalings for amplitudes and their slow 
variation in space and time  were  chosen as   $\hat a,\hat b\sim\gamma^{\frac12},
\dx{\hat a}{x}\sim\dx{\hat a}{t}\sim\gamma$.
 Also, following \cite{Cherep} a stretched coordinate 
 was introduced to resolve the  boundary layer's dynamics.
The time and space variables  have been non-dimensionalized  using
$\om^{-1}$ and $\tk^{-1}$, respectively, as their scales. 

Higher-order corrections to the coefficients of the linear damping 
terms, $\gamma^{\frac32}$ , and of the cubic terms,
 $p_{i}\sqrt{\gamma}$, represent,  respectively,  
dissipation in the viscous boundary layer \cite{Vinals}  
and  nonlinear  interaction between potential wavefield and 
rotational flow  in the boundary layer.
 Due to their cumbersome form, complex-valued 
coefficients $p_i$ are not presented here explicitly.  
  
 Imaginary parts of the coefficients of the cubic terms are
responsible for  saturation of the amplitude growth   by tuning the wave out of the 
resonance  (nonlinear frequency shift).
 Asymptotically small coefficients of the  cubic terms 
 are retained in eqs.\rf{ampl_eq_a},\rf{ampl_eq_b} since nonlinear damping
 can become   important for small supercriticalities \cite{We,Cross_Hohen},  for which
 a  simpler model is derived below. 

A model similar to eqs.\rf{ampl_eq_a},\rf{ampl_eq_b} but 
 without the higher-order damping terms 
 was  first derived using symmetry arguments  and used in \cite{Ezerskii} with 
the coefficients estimated from the experiments.
High-order nonlinear terms with real coefficients were included 
  in \cite{Milner}. However,  as a result of neglecting of  dissipation in the 
 viscous  boundary layer,  the  corresponding coefficients are asymptotically smaller
 ($\sim\gamma$) than in our model. 

 In this paper we consider the case of $p\equiv \Re(p_1+p_2)>0$, in which, as 
we show below,  subcritical solutions  do not exist.  
We also put aside the case of  second-harmonic resonance, $\sg\approx\frac13$,
 in which the coefficients $c_1,p_1$ diverge indicating a  breakdown  of the mode.

Note, that the presence of the   first spatial derivatives  in  eqs.\rf{ampl_eq_a},
\rf{ampl_eq_b} follows  only   from the scaling  
$\partial_x\sim \partial_t\sim\gamma\sim f$ 
which is appropriate for the case under consideration. % at hand. 
Indeed,  according to the linear stability analysis  \cite{Chris,We} the 
neutral curve for  $ f\sim\gamma\ll1$  is defined by the  equation   
$s^2(1-k)^2+ f_c^2= f^2$.
Above the threshold $ f_c=\gamma-\frac12\gamma^{\frac32}$ the bandwidth of unstable 
modes  grows proportionally to $ f$ for a wide range of  supercriticalities $\delta^2=(f-f_c)/f_c$.
Hence, the evolution of the envelopes takes place on  the spatial   scale  
$ f^{-1}$, which makes the spatial derivatives in eqs.\rf{ampl_eq_a},\rf{ampl_eq_b} 
 of the same  asymptotic order as the other terms. A similar ``non-traditionally'' slow 
spatial variation of amplitude takes place in a parametrically 
driven  oscillatory  dissipative system \cite{PRE}. 

We relate  the damping parameter  and  supercriticality  
as  $\gamma=\gamma_1^2\del^2,\,\gamma_1=\order(1)$.   The  expansions 
$ \dx{}{x}=\del^2\dx{}{x_2}+ \del^3\dx{}{x_3}+\ldots,\quad 
\dx{}{t}=\del^2\dx{}{t_2}+ \del^3\dx{}{t_3}+ \ldots,
\quad  \{\hat a,\hat b\} = \del^{\frac32}\{a^{(1)},b^{(1)}\} +  \del^{\frac52}\{a^{(2)},b^{(2)}\} +  \ldots,\quad$ are  then introduced into 
  eqs.\rf{ampl_eq_a},\rf{ampl_eq_b}
and a   hierarchy of  problems is obtained and  solved at each order.
At   $\order(\del^{7/2})$   we  obtain
\be
\dx{a^{(1)}}{t_2}=\dx{a^{(1)}}{x_2}=\dx{b^{(1)}}{t_2}=\dx{b^{(1)}}{x_2}=0,\quad a^{(1)}=ib^{(1)}. 
\label{order1_2}
\ee

At this order we find that the motion has a form of standing waves and   
the dynamics of their envelope evolve on  the  ``slower'' time- and space-scales
 $t_3,x_3\ldots$.

At  $\order(\del^{\frac92})$ we obtain  using eq.\rf{order1_2} two equations
 \ba
\dx{a^{(1)}}{t_3}- s\dx{a^{(1)}}{x_3}= -a^{(2)} + ib^{(2)} + ic |a^{(1)}|^2a^{(1)},\label{order3_2a}
\\
\dx{b^{(1)}}{t_3}+ s\dx{b^{(1)}}{x_3}= -b^{(2)}- ia^{(2)} -ic|a^{(1)}|^2b^{(1)},\label{order3_2b}
\ea
where $c\equiv c_1+c_2$.
Adding eq.\rf{order3_2a} and eq.\rf{order3_2b} multiplied by $i$ 
and using eq.\rf{order1_2} we obtain   $\dx{a^{(1)}}{t_3}=\dx{b^{(1)}}{t_3}=0$.

A solvability condition of eqs.\rf{ampl_eq_a},\rf{ampl_eq_b}
 at order $\del^{\frac{11}{2}}$
 yields an amplitude  equation, which  in terms of 
the unrescaled amplitude $\hat a$  and space and time variables $x,t$  reads 
\be
 \dx{\hat a}{t} = (f- f_c)\hat a -p \sqrt{\gamma}|\hat a|^2\hat  a+\frac{ s^2}{2 f} \ddx{\hat a}{x}+
 \frac{i sc_1}{ f} \dx{(|\hat a|^2\hat a)}{x}+
\frac{i sc_2}{ f} |\hat a|^2\dx{\hat a}{x} - \frac{c^2}{2 f}|\hat a|^4\hat a.
\label{Quintic_unresc}
\ee

This equation  describes the slow evolution of the  envelope of 2D small-amplitude 
 driven standing waves on the surface of a weakly-viscous fluid.
 In what follows   we consider eq.\rf{Quintic_unresc} with 
 periodic boundary conditions  in $0\le x\le L$.
It turns out, that eq.\rf{Quintic_unresc} is also relevant for  a large spectrum of
$\del/f$ \cite{We}.
In the limiting cases of very small and ``large''  supercriticalities 
there are two different  mechanisms of amplitude  saturation in eq.\rf{Quintic_unresc}.
For   supercriticalities $( f- f_c)/ f_c\ll f_c$ the viscous cubic term  is 
dominant and the dynamics has the relaxational form 
 described by the  Ginzburg-Landau equation \cite{Cross_Hohen}.
 In the opposite case of ``large''  supercriticality (or, equivalently, small damping)
 $\gamma\ll( f- f_c)/ f_c\ll1$  saturation is due to nonlinear frequency shift
  provided by the three last terms in eq.\rf{Quintic_unresc}.
  
In this last case, eq.\rf{Quintic_unresc} becomes non-typical, since the ``viscous'' 
cubic term becomes small and can be omitted, and after  appropriate rescaling there is 
only one independent parameter. 
The reason for this strong mathematical
degeneracy is that the coefficients of the cubic terms in
 eqs.\rf{ampl_eq_a},\rf{ampl_eq_b} become purely
imaginary, which is a  result of a   weakly dissipative  character  of the  system. 
In this case the only  mechanism of
amplitude saturation is the nonlinear frequency shift.   
 Along  with the linear dispersion (first derivative terms in  
eqs.\rf{ampl_eq_a},\rf{ampl_eq_b}), it
  leads to  different types of bifurcation away from the trivial state to
 the  monochromatic solution $\hat a=\hat A(t)\exp\{i(k-1)x)\}$
  around the  resonant wavenumber $k=1$: supercritical for $k>1$, 
and subcritical  for $k<1$. 
This implies that the coefficient of the cubic  term in 
the  corresponding amplitude equation for $A$, 
changes its sign at $k=1$, and  thus, in the  vicinity of the minimum the
 saturation is provided by the  quintic term.
 
According to eq.\rf{Quintic_unresc} the driven standing waves have an abnormally 
large amplitude, which in the typical case of $ \gamma\sim( f- f_c)/ f_c\ll1 $ is
 proportional to the quartic root of the  supercriticality.
This peculiar steepness of the off-branching solution constitutes the substantial
difference between patterns formation in a typical strongly-dissipative 
system and a parametrically forced weakly-dissipative oscillatory  system.
 The second important 
difference is a non-traditionally slow spatial variation of the 
envelope,  $\partial_x a \sim ( f- f_c)^{3/4}$, whereas for a dissipative 
system it is   $ \sim ( f- f_c)^{1/2}$ \cite{Cross_Hohen}.

\section{Analytical and numerical study of the model equation} 
We first show  that in the subcritical domain  the equilibrium state 
is absolutely stable.
 To prove this we multiply eq.\rf{Quintic_unresc} by $\hat a^*$ and 
integrate over  $[0,L]$. Adding then the complex conjugate of the latter 
and integrating by parts we obtain 
$$
\frac12\partial_t\int_0^L|\hat a|^2dx=\int_0^L\left(( f- f_c)|\hat a|^2-  
p\sqrt{\gamma}|\hat a|^4-
\frac{1}{2f}\left|is\dx{\hat a}{x}- c|\hat a|^2\hat a\right|^2\right)dx 
$$
that  yields a decay of  perturbations when  $ f< f_c$ and $p>0$.

Assuming now $p>0, f> f_c$ we  rescale eq.\rf{Quintic_unresc} 
using new  variables
$
\tau=tp^2 f\gamma/2c^2,\,\xi= xp f\sqrt{\gamma}/sc,\,
a=\hat a|c|/\sqrt{pf\sqrt{\gamma}}. $
In  a  rescaled form  eq.\rf{Quintic_unresc} reads 
\be
 \dx{a}{\tau} =  \vep a + \ddx{a}{\xi} -2|a|^2a +2i\beta\dx{(|a|^2)}{\xi}a +
2i|a|^2\dx{a}{\xi} - |a|^4a.
 \label{Quintic_resc} 
 \ee
Eq.\rf{Quintic_resc}  contains only two parameters: the effective supercriticality
  $\vep= [( f- f_c)/ f_c](2c^2/\gamma p^2)$,
 and  $\beta=c_1/c$,  whose value  depends  on $\sg$ and varies in 
a wide range.
%, see Eq.\rf{parameters}. 

%The equilibrium state $\bar a=0$ is unstable with  respect to  the perturbations 
%$\sim\exp(iKX),\,|K|<1$.
We now consider the simplest  non-trivial solution of the form  
$a(\xi,\tau)=A(\tau)\exp(iK\xi)$, describing a  standing   monochromatic (MC) wave
with the modified wavenumber   $k= 1+ Kp f\sqrt{\gamma}/{sc}$. 
In this case the time evolution of the  amplitude 
 has the  gradient form
\be
\dx{A}{\tau}=-\dx{F}{A},
\quad\mbox{ where } F=\frac16(K+A^2)^3+\frac12A^4-\frac{\vep}{2}A^2.\label{Gradient}
\ee
Since $F$ is bounded from below for fixed  $K$,  any  MC perturbation 
results either in the trivial solution $a\equiv0$   or in one of the 
stationary solutions 
\be
 a_\pm=\AC_\pm e^{iK\xi},\quad\AC^2_\pm=-K- 1\pm\sqrt{\vep+2K+1}. \label{MSW}
\ee
It  follows from eqs.\rf{Gradient},\rf{MSW} that the solution $a_-$, which 
exists for $K<-1$ and  branches off the trivial solution subcritically,
  is unstable. 
 
The bifurcation diagram of standing MC wave
 is shown in fig.1a for different $\vep$.
It is seen that for  large values of  $\vep$, such that   $\vep f\ll1$ 
(this condition  supports the asymptotic method), a solution with an arbitrarily large 
amplitude is possible. 
If such a solution  materializes  as a result of the  time evolution 
of a  finite perturbation  $|a|\sim 1$, then  this would  mean  the 
failure of the model eq.\rf{Quintic_resc}.
 However, as our numerical study  shows (see below),
there is a mechanism of  the amplitude saturation  owing to  
the   subcritical character of  bifurcation of the large-amplitude solution 
$a_+$ for $K<-1$.
 
We now proceed   to sideband (SB) stability analysis of the MC
solution $a_+$. 
The perturbed solution   is expressed as
\be
 a= e^{iK\xi}(\AC _+ +\al_1 e^{iq\xi+\lambda \tau}+ \al_2 e^{-iq\xi+\lambda^* \tau}).
\label{SB_pert}
\ee
Here $\al_1,\al_2$ are complex amplitudes, $q$ is the perturbation of 
the wave number and  $\lambda$ is the growth rate.
Upon  substitution of eq.\rf{SB_pert} into eq.\rf{Quintic_resc}
 we find 
$$
\lambda=-q^2B(B+K+1)\pm\sqrt{B^2(B+1+K)^2 +q^2(K+B)(K+B+2\beta B)},\quad 
\mbox{ where } B=\AC^2.
$$
The instability is long-wave and the threshold is defined by the minimal  wavenumber 
 allowed by the periodic constraint. 
For an infinite system $q_{min}\to 0$,  and   the threshold  is defined 
by the equation $B=(B+K)(K+2\beta B)$. 
For  positive  $\beta$'s the SB-stability 
domain is adjacent to the  saddle-node bifurcation curve $\vep_{SN}=-2K-1$ 
(curve 1 in  fig.1b),  and shrinks   when $\beta$ increases. 
For negative $\beta$'s the SB-stability domain widens and moves towards
the center of the linear instability domain  bounded by the neutral   
 curve $\vep=K^2 $ (dotted curve in  fig.1b).

We now  show an  existence  of  a stationary    non-MC 
 solution of the  form  $a=A(\xi)\exp(i\th(\xi))$. For this   
 we separate eq.\rf{Quintic_resc} into real and imaginary parts to obtain
\ba
\vep A + \ddx{A}{\xi} -2A^3-  (\dx{\th}{\xi})^2A 
 -2 A^3\dx{\th}{\xi} - A^5=0, \label{separa}\\
2\dx{A}{\xi}\dx{\th}{\xi}+ A\ddx{\th}{\xi} +2(2\beta+1)A^2\dx{A}{\xi}=0.\label{separb}
\ea
Eq.\rf{separb} can be integrated allowing an  elimination 
of  the phase $\th$ from  eq.\rf{separa}:
\be
\ddx{A}{\xi}= -\dx{U}{A},\quad \mbox{ where }
 U= \frac{C^2}{2A^2} +\frac{A^2}{2}(\vep+2C\beta-C) -\frac12A^4
 - \frac{A^6}{24}(1-2\beta)^2.
 \label{particle}
\ee
This equation   describes a particle  motion in a potential 
 field,  where  $\xi$ is interpreted as  time.  
The potential $U$ has a local minimum at  $A=\AC_1$.  
A  periodic motion of the particle  around the minimum   corresponds to
 a non-MC  solution periodic in space.
Two integration constants can be  determined from the  periodicity 
conditions  for the amplitude $A$  and the  local wavenumber $\partial_\xi\theta$. 
 In the presence of the  periodic constraint one finds 
 a countable   family of periodic solutions, which are spatially non-MC. 
In an unbounded system  (in the absence of  periodic constraint)    
 the continuum of  solutions exists, including a stationary solitary wave
 corresponding to a homoclinic orbit in the phase plane 
of eq.\rf{particle}.

We have also investigated  numerically  Eq.\rf{Quintic_resc} 
along with periodic boundary 
conditions  to study  more complex  non-stationary regimes.
We used a  pseudo-spectral method with a second-order 
time-marching and up to $256$ spatial Fourier modes.   
In a full agreement with our stability analysis, 
the SB-unstable MC solution  breaks down if it is 
initially  perturbed by a sideband perturbation. 
The resulting regime depends on whether for a given  $\vep$,  all
 SB-stable solutions are deeply subcritical ({\it e.g.} $\beta=2,\vep=8$ in fig.1b)  or 
 there exist  SB-stable solutions in a supercritical domain 
({\it e.g.} $\beta=-1$ in fig.1b).

 In the first case   the breakdown of an  unstable solution leads 
(via a  phase-slip in the underlying standing waves)  to  onset of the 
stationary  MC  waves  with a different  wavenumber from the 
SB-stability domain (see fig.2a).
This is a typical behavior for a negative or slightly positive 
$\beta$, for which the domains  of SB-stability  and instability of the
 trivial solution overlap for moderate $\vep$.  
Non-MC solutions studied above analytically, were not detected 
 in our numerical  simulations. 
They  seem  to  belong  to unstable manifolds 
 separating  the basins of  attraction of MC solutions.

For  positive $\beta$ and  large values of  $\vep$,
 for  which SB-stable solutions are
 in the  subcritical domain (see the  horizontal dashed line in fig.1b),
   the instability of MC waves typically results
in  onset of an irregular regime. 
The time-evolution of the Fourier spectrum for  
$\beta=1,\del=8$  is shown in fig.2b.
This irregular behavior was checked using a  refined numerical scheme 
during a rather  long calculation  ($\tau=10^4$), in which  
 it remained qualitatively the same.  

 In our opinion this irregular behavior is a result of two 
factors: (i) SB instability of MC solutions with moderate $K$,
 and (ii)  subcritical character of bifurcation of 
SB-stable MC solutions, which  does not allow  the perturbations with  
large negative $K$  to attain their  stationary values.

To summarize our results, a new evolution
equation describing the weakly 
supercritical regimes of driven surface waves
was derived and studied.
 Its quintic form results in solutions of 
a non-traditionally  large amplitude.
 The sideband instability of a MC wave  leads to the onset of either 
a MC wave with a different wavenumber or an  irregular motion.   \medskip

\stars

We thank A.A.Nepomnyashchy for helpful discussions.
This work was  supported by the Center for Absorption in Science,
Ministry of Immigrant Absorption, State of Israel (to I. K.), and by 
Y.Winograd Chair of Fluid Mechanics and Heat Transfer at Technion.
A.O. was partially supported by the Fund for the Promotion of Research
at the Technion.

\begin{figure}
\epsfxsize=14.5cm
\epsffile{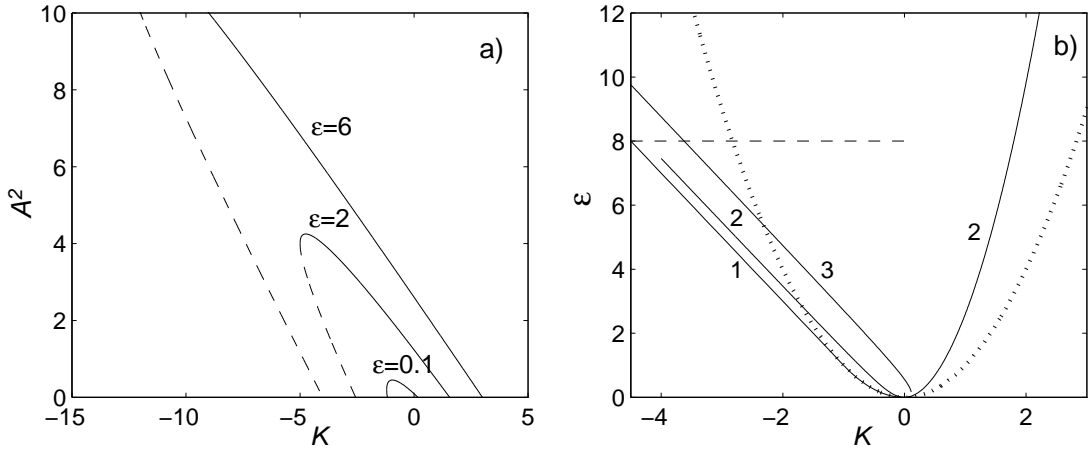}
\caption{Bifurcation diagram (a) and stability map (b) of the MC solutions.
In (a) stable and unstable branches are shown by solid and dashed curves, respectively.
The dotted  curve in (b)  denotes the linear stability threshold 
of the trivial state.  
The domain of  SB-stability of the 
 MC solution for $\beta=-1$ is bounded from below by curve 2, 
whereas for $\beta=2$ it is contained  between the
 saddle-node curve 1 and  curve 3.}
\end{figure}

\begin{figure}
\epsfxsize=14.5cm
\epsffile{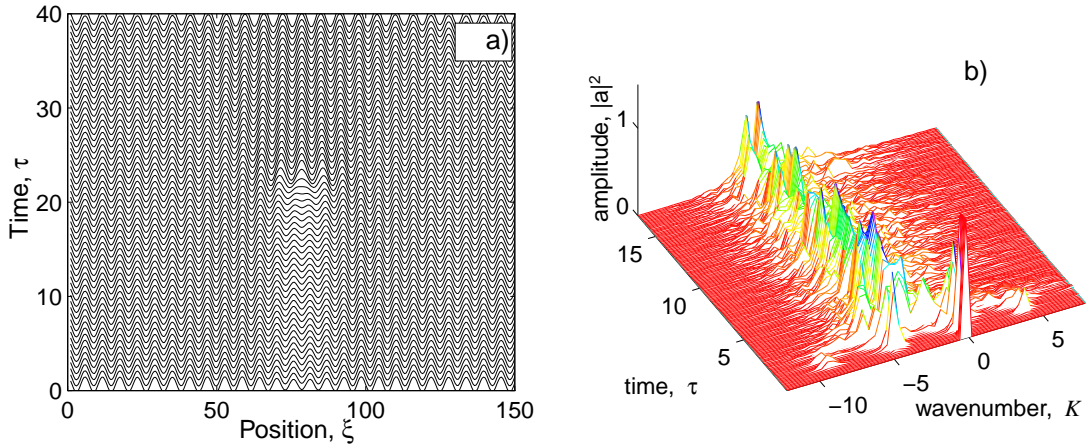}
\caption{SB-instability of the resonant $(K=0)$ standing MC wave
 leading for $\beta=-1$ (a)  to onset of  MC wave 
with  a different wavenumber, and   for $\beta=2$ (b)  to a persistent 
irregular pulsing (shown in Fourier space). The  bandwidth of the pulsing is
 shown in fig.1b by the horizontal crossing line.  
 }
\end{figure}

\end{document}